\documentclass[12pt]{article}
\usepackage{amssymb,amsmath}

\let\OLDthebibliography\thebibliography
\renewcommand\thebibliography[1]{
  \OLDthebibliography{#1}
  \setlength{\parskip}{0pt}
  \setlength{\itemsep}{0pt plus 0.3ex}
}

\newcommand{\gam}{\gamma}

\newcommand{\tl}{\tilde}
\newcommand{\ul}{\underline}

\newcommand{\be}{\begin{equation}}
\newcommand{\bear}{\begin{eqnarray}}
\newcommand{\ear}{\end{eqnarray}}
\newcommand{\ee}{\end{equation}}
\newcommand{\lbl}{\label}
\newcommand{\bi}{\bibitem}
\newcommand{\ci}{\cite}

\newcommand{\vs}{\vspace}

\begin{document}

\

\baselineskip .7cm 

\vs{8mm}

\begin{center}

{\LARGE \bf Geometric Spinors, Generalized Dirac Equation and Mirror Particles }

\vs{3mm}

Matej Pav\v si\v c

Jo\v zef Stefan Institute, Jamova 39,
1000 Ljubljana, Slovenia

e-mail: matej.pavsic@ijs.si

\vs{6mm}

{\bf Abstract}
\end{center}

\baselineskip .4cm 

{\footnotesize
It is shown that since the geometric spinors are elements of Clifford
algebras, they must have the same transformation properties as any other
Clifford number. In general, a Clifford number $\Phi$ transforms into
a new Clifford number $\Phi'$ according to
$\Phi  \to \Phi ' = {\rm{R}}\,\Phi \,{\rm{S}}$, i.e.,  by the multiplication
from the left and from the right by two Clifford numbers R and S. We study
the case of $Cl(1,3)$, which is the Clifford algebra of the Minkowski spacetime.
Depending on choice of R and S, there are various possibilities, including
the transformations of vectors into 3-vectors, and the transformations of
the spinors of one minimal left ideal of $Cl(1,3)$ into another minimal
left ideal. This, among others, has implications for understanding the observed
non-conservation of parity.
}

\baselineskip .5cm

\section{Introduction}

We will follow the approach\,\ci{SpinorFock}--\ci{PavsicSpinorInverse}
 in which spinors are constructed in terms
of nilpotents formed from the spacetime basis vectors represented
as generators of the Clifford algebra

\be
\begin{array}{l}
 \gamma _a \,\cdot\,\gamma _b \, \equiv \,\frac{1}{2}\,(\gamma _a \gamma _b
   + \gamma _b \gamma _a )\, = \eta _{ab}  \\ 
 \gamma _a  \wedge \gamma _b \, \equiv \,\frac{1}{2}\,(\gamma _a \gamma _b 
  - \gamma _b \gamma _a ). \\ 
 \end{array} \lbl{1}
\ee
     
The inner, symmetric, product of basis vectors
gives the metric.  The outer, antisymmetric, product of basis vectors
gives a basis bivector.

The generic Clifford number is
\be
\Phi  = \varphi ^A \gamma _A .
\lbl{2}
\ee
where
$\gamma _A  \equiv \gamma _{a_1 a_2 ...a_r }  \equiv 
\gamma _{a_1 }  \wedge \gamma _{a_2 }  \wedge ... \wedge \gamma _{a_r }, 
~~~r=0,1,2,3,4$.

Spinors are particular Clifford numbers, 
$\Psi  = \psi ^\alpha  \xi _\alpha$, 
where $\xi_\alpha$ are spinor basis elements, composed from $\gam_A$. 
We will consider transformation properties of Clifford numbers.

 In general, a Clifford number transforms according to
\be
\Phi  \to \Phi ' = {\rm{R}}\,\Phi \,{\rm{S}}.
\lbl{4}
\ee
Here $R$ and $S$ are Clifford numbers, e.g.,
${\rm{R}} = {\rm{e}}^{\frac{1}{2}\alpha ^A \gamma _A }$, ${\rm{S}}
 = {\rm{e}}^{\frac{1}{2}\beta ^A \gamma _A }$.

In particular, if $S=1$, we have 
\be
\Phi  \to \Phi ' = {\rm{R}}\,\Phi .
\lbl{6}
\ee
As an example, let us consider the case
\be
{\rm{R}} = {\rm{e}}^{\frac{1}{2}\alpha \gamma _1 \gamma _2 } 
 = {\rm{cos}}\,\frac{\alpha }{2} + \gamma _1 \gamma _2 \,{\rm{sin}}\,
 \frac{\alpha }{2}\:,\,\,\,\,\,\,\,\,\,\,\,\,\,\,{\rm{S}} 
 = {\rm{e}}^{\frac{1}{2}\beta \gamma _1 \gamma _2 } 
  = {\rm{cos}}\,\frac{\beta }{2} + \gamma _1 \gamma _2 \,{\rm{sin}}\,\frac{\beta }{2}
\lbl{7} \ee
and examine\,\ci{PavsicE8}, how various Clifford numbers,
\be
X = X^C \gamma _C, 
\ee
transform under (\ref{4}), which now reads:
\be
X \to X' = {\rm{R}}\,X\,{\rm{S}} .
\lbl{8} \ee

 (i) If  $X = X^1 \gamma _1  + X^2 \gamma _2$ then
\be
X' = X^1 \left( {\gamma _1 \,{\rm{cos}}\,\frac{{\alpha  - \beta }}{2}\,
 + \gamma _2 \,{\rm{sin}}\,\frac{{\alpha  - \beta }}{2}} \right) 
 + X^2 \left( { - \gamma _1 \,{\rm{sin}}\,\frac{{\alpha  - \beta }}{2}\, 
 + \gamma _2 \,{\rm{cos}}\,\frac{{\alpha  - \beta }}{2}} \right).
\lbl{10} \ee

(ii) If $X = X^3 \gamma _3  + X^{123} \,\gamma _{123}$ then 
\be
X' = X^3 \left( {\gamma _3 \,{\rm{cos}}\,\frac{{\alpha  + \beta }}{2}\,
 + \gamma _{123} \,{\rm{sin}}\,\frac{{\alpha  + \beta }}{2}} \right)
 + X^{123} \left( { - \gamma _3 \,{\rm{sin}}\,\frac{{\alpha  + \beta }}{2}\,
  + \gamma _{123} \,{\rm{cos}}\,\frac{{\alpha  + \beta }}{2}} \right) .
\lbl{10a} \ee

(iii)  If $X = s\, {\ul 1} + X^{12} \,\gamma _{12}$, then
\be
X' = s \left ( {\ul 1} \,{\rm cos}\,\frac{\alpha  + \beta}{2}\, 
+ \gamma_{12} \,{\rm sin}\,\frac{\alpha  + \beta}{2} \right )
 + X^{12} \left( { - {\ul 1} \,{\rm{sin}}\,\frac{{\alpha  + \beta }}{2}\,
  + \gamma _{12} \,{\rm{cos}}\,\frac{{\alpha  + \beta }}{2}} \right) .
\lbl{10b}
\ee

(iv) If $X = \tilde X^1 \,\gamma _5 \gamma _1  
+ \tilde X^2 \,\gamma _5 \gamma _2$, then
\be
X' = \tilde X^1 \left( {\gamma _5 \gamma _1 \,{\rm{cos}}\,
\frac{{\alpha  - \beta }}{2}\, + \gamma _5 \gamma _2 
\,{\rm{sin}}\,\frac{{\alpha  - \beta }}{2}} \right) + \tilde X^2 
\left( { - \gamma _5 \gamma _1 \,{\rm{sin}}\,\frac{{\alpha  - \beta }}{2}\,
 + \gamma _5 \gamma _2 \,{\rm{cos}}\,\frac{{\alpha  - \beta }}{2}} \right) .
\lbl{14} \ee

Usual rotations of vectors or pseudovectors are reproduced,
if the angle $\beta$  for the right transformation is equal to minus angle          
for the left transformation, i.e., if $\beta \,\, = \,\, - \alpha$. 
Then all other transformations which mix the grade vanish. But in general, if
$\beta \neq \alpha$, the transformation (\ref{8}) mixes the grade.

\section{Clifford algebra and spinors in Minkowski space}

Let us introduce a new basis, called the Witt basis,
\be
\begin{array}{l}
 \theta _1  = \frac{1}{2}(\gamma _0  + \gamma _3 )\,,\theta _2  
 = \frac{1}{2}(\gamma _1  + i\gamma _2 )\,, \\ 
 \bar \theta _1  = \frac{1}{2}(\gamma _0  - \gamma _3 )\,,\bar \theta _2 
  = \frac{1}{2}(\gamma _1  - i\gamma _2 ) , \\ 
 \end{array}
\lbl{16} \ee
where \be
\gamma _a  = (\gamma _0 ,\gamma _1 ,\gamma _2 ,\gamma _3 ) .
\lbl{17} \ee
The new basis vectors satisfy
\be
\,\{ \theta _a ,\bar \theta _b \}  = \eta _{ab} ,~~~~\{ \theta _a ,\theta _b \}  = 0,
~~~~\{ \bar \theta _a ,\bar \theta _b \}  = 0 ,
\lbl{18} \ee
which are fermionic anticommutation relations. We now observe that the product
\be
f = \bar \theta _1 \bar \theta _2 
\lbl{19} \ee
satisfies
\be
\bar \theta _a \,f = 0\,,\,\,\,\,\,\,\,\,\,\,\,\,\,\,\,\,\,a = 1,2 .
\lbl{20} \ee
Here $f$ can be interpreted as a `vacuum',
and $\bar \theta _a$ can be interpreted as operators
that annihilate  $f$.   

An object constructed as a superposition
\be
\Psi  = (\psi ^0  + \psi ^1 \theta _1  + \psi ^2 \theta _2  + \psi ^{12} \theta _1 \theta _2 )f
\lbl{23} \ee
is a 4-component spinor. It is convenient to change the notation:
\be
\Psi  = (\psi ^1  + \psi ^2 \theta _1 \theta _2  + \psi ^3 \theta _1 
 + \psi ^4 \theta _2 )f\, = \psi ^\alpha  \xi _\alpha  \,,\,\,
 \,\,\,\,\,\,\,\,\,\,\,\,\alpha  = 1,2,3,4
\lbl{24} \ee
where $\xi_\alpha$ is the spinor basis.

The even part of the above expression is a left handed spinor
\be
\Psi _L  = (\psi ^1  + \psi ^2 \theta _1 \theta _2 )\,\bar \theta _1 \bar \theta _2 ,
\lbl{25} \ee
whereas the odd part is a right handed spinor
\be
\Psi _R  = (\psi ^3 \theta _1  + \psi ^4 \theta _2 )\bar \theta _1 \bar \theta _2 . 
\lbl{26} \ee
We can verify that the following relations are satisfied:
\be
i\gamma _5 \Psi _L  =  - \Psi _L ,~~~~~~~
i\gamma _5 \Psi _R  = \Psi _R 
\lbl{28} \ee
Under the transformations
\be
\Psi  \to \Psi ' = {\rm{R}}\Psi \,,
\lbl{29} \ee
where \be
{\rm{R}} = \exp [\frac{1}{2}\gamma _{a_1 } \gamma _{a_2 } \varphi ] ,
\lbl{30} \ee
the Clifford number $\Psi$ transforms as a spinor.

As an example let us consider the case
\be
{\rm{R}} = {\rm{e}}^{\frac{1}{2}\gamma _1 \gamma _2 \varphi }  
= {\rm{cos}}\,\frac{\varphi }{2} + \gamma _1 \gamma _2 \,{\rm{sin}}\,\frac{\varphi }{2}.
\lbl{31} \ee
Then we have
\be
\Psi  \to \Psi ' = {\rm{R}}\Psi \, = \left( {{\rm{e}}^{\frac{{i\phi }}{2}}
 \psi ^1  + {\rm{e}}^{ - \frac{{i\phi }}{2}} \psi ^2 \theta _1 \theta _2 
  + {\rm{e}}^{\frac{{i\phi }}{2}} \psi ^3 \theta _1  
  + {\rm{e}}^{ - \frac{{i\phi }}{2}} \psi ^4 \theta _2 } \right)f .
\lbl{32} \ee
This is the well-known transformation of a 4-component spinor.

\subsection{Four independent spinors}

There exist four different possible
vacua\,\ci{PavsicSpinorInverse,PavsicOrthoSympl,BudinichFock}:
\be
f_1  = \bar \theta _1 \bar \theta _2 \,,\,\,\,\,\,\,\,\,\,f_2 
 = \theta _1 \theta _2 \,,\,\,\,\,\,\,\,\,\,f_3 
  = \theta _1 \bar \theta _2 \,,\,\,\,\,\,\,f_3  = \bar \theta _1 \theta _2 
\lbl{33} \ee
to which there correspond four different kinds of spinors:
\be
\begin{array}{l}
 \Psi ^1  = (\psi ^{11}  + \psi ^{21} \theta _1 \theta _2 
  + \psi ^{31} \theta _1  + \psi ^{41} \theta _2 )f_1  \\ 
 \Psi ^2  = (\psi ^{12}  + \psi ^{22} \bar \theta _1 \bar \theta _2  
 + \psi ^{32} \bar \theta _1  + \psi ^{42} \bar \theta _2 )f_2  \\ 
 \Psi ^3  = (\psi ^{13} \bar \theta _1  + \psi ^{23} \theta _2 
  + \psi ^{33}  + \psi ^{43} \bar \theta _1 \theta _2 )f_3  \\ 
 \Psi ^4  = (\psi ^{14} \theta _1  + \psi ^{24} \bar \theta _2 
  + \psi ^{34}  + \psi ^{44} \theta _1 \bar \theta _2 )f_4 . 
 \end{array}
\lbl{34} \ee
Each of those spinors lives in a different minimal left ideal
of $Cl(1,3)$, or in general, of its complexified version if we assume complex
$\psi^{\alpha i}$.
  
An arbitrary element of  $Cl(1,3)$ is the sum:
\be
\Phi  = \Psi ^1  + \Psi ^2  + \Psi ^3  + \Psi ^4 
 = \psi ^{\alpha i} \xi _{\alpha i}  \equiv \psi ^{\tilde A} \xi _{\tilde A} ,
\lbl{35} \ee
where
\be
\xi _{\tilde A}  \equiv \xi _{\alpha i} 
 = \{ f_1 ,\,\,\theta _1 \theta _2 f_1 ,...,\theta _1 f_4 ,\,\,
 \bar \theta _2 f_4 ,\,\,f_4 ,\,\,\bar \theta _1 \theta _2 f_4 \} ,
\lbl{36} \ee
is a spinor basis of $Cl(1,3)$. Here $\Phi$ is a generalized spinor.
  
In matrix notation we have
\be
\psi ^{\alpha i}  = \left( {\begin{array}{*{20}c}
   {\psi ^{11} } & {\psi ^{12} } & {\psi ^{13} } & {\psi ^{14} }  \\
   {\psi ^{21} } & {\psi ^{22} } & {\psi ^{23} } & {\psi ^{24} }  \\
   {\psi ^{31} } & {\psi ^{32} } & {\psi ^{33} } & {\psi ^{34} }  \\
   {\psi ^{41} } & {\psi ^{42} } & {\psi ^{43} } & {\psi ^{44} }  \\
\end{array}} \right)\,,\,\,\,\,\,\,\,\,\,\xi _{\tilde A}  
\equiv \xi _{\alpha i}  = \left( {\begin{array}{*{20}c}
   {f_1 } & {f_2 } & {\bar \theta _1 f_3 } & {\theta _1 f_4 }  \\
   {\theta _1 \theta _2 f_1 } & {\bar \theta _1 \bar \theta _2 f_2 } 
   & {\theta _2 f_3 } & {\bar \theta _2 f_4 }  \\
   {\theta _1 f_1 } & {\bar \theta _1 f_2 } & {f_3 } & {f_4 }  \\
   {\theta _2 f_1 } & {\bar \theta _2 f_2 } 
   & {\bar \theta _1 \theta _2 f_3 } & {\theta _1 \bar \theta _2 f_4 }  \\
\end{array}} \right) .
\lbl{37} \ee
Here, for instance, the second column in the left matrix contains
the components of the spinor of the second left ideal. Similarly,
the second column in the right matrix contains the basis elements
of the second left ideal. 

A general transformation is
\be
\Phi  = \psi ^{\tilde A} \xi _{\tilde A}  \to \Phi ' 
= {\rm{R}}\,\Phi \,{\rm{S}} = \,\psi ^{\tilde A} \xi '_{\tilde A} 
 = \psi ^A {L_{\tl A}}^{\tl B} \xi _B  = \psi '^{\tl B} \xi_{\tl B} 
\lbl{38} \ee
where \be
\xi '_{\tilde A}  = \,{\rm{R}}\xi _{\tilde A} {\rm{S}} 
= {L_{\tilde A}}^{\tilde B} \xi _{\tilde B} \,,\,\,\,\,\,\,\,\,\,\,\,\,
\psi '^{\tilde B}  = \psi ^{\tilde A} {L_{\tilde A}}^{\tilde B} .
\lbl{39} \ee
This is an active transformation, because it changes an object
$\Phi$ into another object $\Phi'$.

The transformation from the left,
\be
\Phi ' = {\rm{R}}\,\Phi  ,
\lbl{40} \ee
reshuffles the components within each left ideal,
whereas the transformation from the right,
\be
\Phi ' = \Phi \,{\rm{S}} ,
\lbl{41} \ee
reshuffles the left ideals.

\section{Behavior of spinors under Lorentz transformations}

Let us consider the following transformation
of the basis vectors
\be
\gamma _a  \to \gamma '_a  = {\rm{R}}\,\gamma _a {\rm{R}}^{ - 1}~,~~~~~~a=0,1,2,3, 
\lbl{42} \ee
where R is a proper or improper
Lorentz transformation. A generalized spinor, $\Phi  \in Cl(1,3)$,
composed of  $\gam_a$, then transforms according to      
\be
\Phi  = \psi ^{\tilde A} \xi _{\tilde A}  \to \Phi ' 
= \,\psi ^{\tilde A} \xi '_{\tilde A}  
= \psi ^A {\rm{R}}\,\xi _B {\rm{R}}^{ - 1}  = {\rm{R}}\,\Phi \,{\rm{R}}^{ - 1} .
\lbl{43} \ee
The transformation (\ref{42}) of the basis vectors has for a
consequence that the object $\Phi$  does not transform
only from the right, but also from the left. This had led
Piazzese to the conclusion that spinors cannot be interpreted
as the minimal ideals of Clifford algebras\,\ci{Piazzese}.

But if the reference frame transforms as
\be
\gamma _a  \to \gamma '_a  = {\rm{R}}\,\gamma _a ,
\lbl{44} \ee
then \be
\Phi  = \psi ^{\tilde A} \xi _{\tilde A}  \to \Phi ' 
= \,\psi ^{\tilde A} \xi '_{\tilde A} 
 = \psi ^{\tilde A} {\rm{R}}\,\xi _{\tilde B}
  = {\rm{R}}\,\Phi .
\lbl{45} \ee
This is a transformation of a spinor. Therefore, the description of spinors
in terms of ideals is consistent.

As we have seen in Sec.1, the transformation (\ref{44}) is also a possible
transformation within  a Clifford algebra. It is a transformation that
changes the grade of a basis element.
Usually, we do not consider such transformations of basis vectors.
Usually reference frames are ``rotated'' (Lorentz rotated)
according to
\be 
\gamma _a  \to \gamma '_a  = {\rm{R}}\,\gamma _a {\rm{R}}^{ - 1} 
 = {L_a}^b \gamma _b  ,
\lbl{46} \ee
where ${L_a}^b$ is a proper or improper Lorentz transformation.
Therefore, a ``rotated'' observer sees  (generalized) spinors
transformed according to
\be
\Phi  \to \Phi ' = \,{\rm{R}}\,\Phi \,{\rm{R}}^{ - 1} .
\lbl{47} \ee
With respect to a new reference
frame, the object $\Phi  = \psi ^{\tilde A} \xi _{\tilde A} $ is expanded
according to
\be
\Phi  = \psi '^{\tilde A} \xi '_{\tilde A} ,
\lbl{48} \ee
where \be
\psi '^{\tilde A}  = \psi ^{\tilde B} {(L^{ - 1} )_{\tilde B}}^{\tilde A} . 
\lbl{49} \ee
Recall that $
 \alpha ,\beta  = 1,2,3,4$, and $i,j = 1,2,3,4$.
The corresponding matrix  $\psi^{\alpha i}$      transforms
from the left and from the right.

If the observer, together with the reference
frame, starts to rotate, then after having
exhibited the  $\varphi=2 \pi$ turn, he observes
the same spinor $\Psi$, as he did at  $\varphi=0$.
The sign of the spinor did not change, because this was just a passive
transformation, so that the same (unchanged) objects was observed from
the transformed (rotated) references frames at different angles $\varphi$.
In the new reference frame the object was observed to be transformed according
to $\Psi'=R\Psi R^{-1}$. There must also exist the corresponding {\it active}
transformation such that in a fixed reference frame the spinor transforms as
$\Psi'=R\Psi R^{-1}$.

\subsection{Examples}

\subsubsection{Rotation}

Let us consider the following rotation:
\be
\begin{array}{l}
 \gamma _0  \to \gamma _0 \,,\,\,\,\,\,\gamma _1  \to \gamma _1 \,,\,
 \,\,\,\,\gamma _2  \to \gamma _2 \cos \,\vartheta \,
  + \gamma _3 \,\sin \,\vartheta  \\ 
 ~~~~~~~~~~~~~~~~~~~~~~~~~~~~~~
 \gamma _3  \to  - \gamma _2 \sin \vartheta  + \gamma _3 \cos \vartheta . 
 \end{array}
\lbl{50} \ee
In the case $\vartheta = \pi$, we have
\be
\gamma _0  \to \gamma _0 \,,\,\,\,\,\,\gamma _1  \to \gamma _1 \,,\,\,
\,\,\,\gamma _2  \to  - \gamma _2 \,,\,\,\,\,\,\gamma _3  \to  - \gamma _3 .
\lbl{51} \ee
The Witt basis then transforms as
\be
 \theta _1  \to \bar \theta _1\, , ~~~~ 
 \theta _2  \to \bar \theta _2 \, ,~~~~ 
 \bar \theta _1  \to \theta _1 \, ,~~~~ 
 \bar \theta _2  \to \theta _2 .
\lbl{52} \ee
A consequence is that, e.g.,
a spinor of the first left ideal transforms as
\be
(\psi ^{11}  + \psi ^{21} \theta _1 \theta _2  + \psi ^{31} \theta _1
  + \psi ^{41} \theta _2 )\,\bar \theta _1 \bar \theta _2 \,\, 
  \to (\psi ^{11}  + \psi ^{21} \bar \theta _1 \bar \theta _2 \, 
  + \psi ^{31} \bar \theta _1  + \psi ^{41} \bar \theta _2 )\,
  \theta _1 \theta _2 .
\lbl{53} \ee
By inspecting the latter relation and taking into account Eqs.(\ref{34}),
(\ref{25}),(\ref{26}),  we see that 
a left handed spinor of the {\it first ideal} transforms into
a left handed spinor of the {\it second ideal}. Similarly, 
a right handed spinor of the first ideal transforms into
a right handed spinor of the second ideal.

In general, under the $\vartheta=\pi$  rotation in the $(\gam_2,\gam_3)$ plane,
 a generalized spinor
\be
\begin{array}{l}
 \Phi \,\, = \,\,(\psi ^{11}  + \psi ^{21} \theta _1 \theta _2 
  + \psi ^{31} \theta _1  + \psi ^{41} \theta _2 )\bar \theta _1 \bar \theta _2  \\ 
 \,\,\,\,\,\,~~ + (\psi ^{12}  + \psi ^{22} \bar \theta _1 \bar \theta _2 
  + \psi ^{32} \bar \theta _1  + \psi ^{42} \bar \theta _2 )\theta _1 \theta _2  \\ 
 \,\,\,\,\,\,~~ + (\psi ^{13} \bar \theta _1  + \psi ^{23} \theta _2  
 + \psi ^{33}  + \psi ^{43} \bar \theta _1 \theta _2 )\theta _1 \bar \theta _2  \\ 
 \,\,\,\,\,\,~~ + (\psi ^{14} \theta _1  + \psi ^{24} \bar \theta _2  
 + \psi ^{34}  + \psi ^{44} \theta _1 \bar \theta _2 )\bar \theta _1 \theta _2  \\ 
 \end{array}
\lbl{54} \ee
transforms into
\be
\begin{array}{l}
 \Phi '\,\, = \,\,(\psi ^{11}  + \psi ^{21} \bar \theta _1 \bar \theta _2 
  + \psi ^{31} \bar \theta _1  + \psi ^{41} \bar \theta _2 )\theta _1 \theta _2  \\ 
 \,\,\,\,\,\,~~ + (\psi ^{12}  + \psi ^{22} \theta _1 \theta _2  + \psi ^{32} \theta _1 
  + \psi ^{42} \theta _2 )\bar \theta _1 \bar \theta _2  \\ 
 \,\,\,\,\,\,~~ + (\psi ^{13} \theta _1  + \psi ^{23} \bar \theta _2 
  + \psi ^{33}  + \psi ^{43} \theta _1 \bar \theta _2 )\bar \theta _1 \theta _2  \\ 
 \,\,\,\,\,\,~~ + (\psi ^{14} \bar \theta _1  + \psi ^{24} \theta _2  
 + \psi ^{34}  + \psi ^{44} \bar \theta _1 \theta _2 )\theta _1 \bar \theta _2  \\ 
 \end{array} .
\lbl{55} \ee
The matrix of components
\be
\psi ^{\alpha i}  = \left( {\begin{array}{*{20}c}
   {\psi ^{11} } & {\psi ^{12} } & {\psi ^{13} } & {\psi ^{14} }  \\
   {\psi ^{21} } & {\psi ^{22} } & {\psi ^{23} } & {\psi ^{24} }  \\
   {\psi ^{31} } & {\psi ^{32} } & {\psi ^{33} } & {\psi ^{34} }  \\
   {\psi ^{41} } & {\psi ^{42} } & {\psi ^{43} } & {\psi ^{44} }  \\
\end{array}} \right) ~~~
{\rm transforms~ into}~~
\psi '^{\alpha i}  = \left( {\begin{array}{*{20}c}
   {\psi ^{12} } & {\psi ^{11} } & {\psi ^{14} } & {\psi ^{13} }  \\
   {\psi ^{22} } & {\psi ^{21} } & {\psi ^{24} } & {\psi ^{23} }  \\
   {\psi ^{32} } & {\psi ^{31} } & {\psi ^{34} } & {\psi ^{33} }  \\
   {\psi ^{42} } & {\psi ^{41} } & {\psi ^{44} } & {\psi ^{43} }  
\end{array}} \right) .
\lbl{57} \ee
We see that in the transformed matrix, the first and the second column are interchanged.
Similarly, also the third and forth column are interchanged. Different columns represent
different left minimal ideals of $Cl(1,3)$, and thus different spinors.

Let us now focus our attention on the spinor basis states of the first and second ideal:
\be
 \xi _{11}  = \bar \theta _1 \bar \theta _2 \, ,~~~~
 \xi _{21}  = \theta _1 \theta _2 \bar \theta _1 \bar \theta _2 \, , ~~~~ 
 \xi _{12}  = \theta _1 \theta _2 \, ,~~~~
 \xi _{22}  = \bar \theta _1 \bar \theta _2 \theta _1 \theta _2  . 
\lbl{58} \ee
which span the left handed part of the 4-component spinor
(see Eqs.(\ref{25},\ref{26})).

Under the $\vartheta = \pi$ rotation (\ref{51}), (\ref{52}), we have
\be
 \xi _{11}  \to \xi _{12} \,,~~~~
 \xi _{21}  \to \xi _{22} \, ,~~~~ 
 \xi _{12}  \to \xi _{11} \,,~~~~
 \xi _{22}  \to \xi _{21} ,
\lbl{61} \ee
which means that the spin 1/2 state of the 1st ideal transforms into the spin           
state of the 2nd ideal, and vice versa. The above states are eigenvalues of the
spin operator, $ - \frac{i}{2}\,\gamma _1 \gamma _2 $,
\be
 - \frac{i}{2}\gamma _1 \gamma _2 \,\xi _{11}  = \frac{1}{2}\xi _{11} \,,\,\,\
 ,\,\,\,\,\,\,\,\, - \frac{i}{2}\gamma _1 \gamma _2 \,\xi _{21} 
  =  - \frac{1}{2}\xi _{21} \, ,
\lbl{63} \ee
\be
 - \frac{i}{2}\gamma _1 \gamma _2 \,\xi _{12}  =  - \frac{1}{2}\xi _{12} \,,\,
 \,\,\,\,\,\,\,\, - \frac{i}{2}\gamma _1 \gamma _2 \,\xi _{22}  
 = \frac{1}{2}\xi _{22} \, . 
\lbl{64} \ee
Let us now introduce the new basis states
\be
\begin{array}{l}
 \xi _{1/2}^1  = \frac{1}{{\sqrt 2 }}(\xi _{11}  + \xi _{22} )\,,\,
 \,\,\,\,\,\,\,\xi _{1/2}^2  = \frac{1}{{\sqrt 2 }}(\xi _{11}  - \xi _{22} ) \, ,\\ 
 \,\xi _{ - 1/2}^1  = \frac{1}{{\sqrt 2 }}(\xi _{21}  + \xi _{12} )\,,\,\
 ,\,\,\xi _{ - 1/2}^2  = \frac{1}{{\sqrt 2 }}(\xi _{21}  - \xi _{12} ) \, . \\ 
 \end{array}
\lbl{65} \ee
which are superpositions of the states of the 1st and the 2nd ideal.
Under the rotation (\ref{51}),(\ref{52}) we have
\be
\begin{array}{l}
 \xi _{1/2}^1  \to \frac{1}{{\sqrt 2 }}(\xi _{12}  + \xi _{21} ) 
 = \xi _{ - 1/2}^1 \, , \\ 
 \,\xi _{ - 1/2}^1  \to \frac{1}{{\sqrt 2 }}(\xi _{22}  + \xi _{11} )
  = \,\xi _{1/2}^1 \, ,  \\ 
 \end{array}
\lbl{66} \ee
\be
\begin{array}{l}
 \xi _{1/2}^2  \to \frac{1}{{\sqrt 2 }}(\xi _{12}  - \xi _{21} ) 
 =  - \xi _{ - 1/2}^2 , \\ 
 \,\xi _{ - 1/2}^2  \to \frac{1}{{\sqrt 2 }}(\xi _{22}  - \xi _{11} )
  = \, - \xi _{1/2}^2 \, . \\ 
 \end{array}
\lbl{67} \ee
These states also have definite spin projection:
\be
 - \frac{i}{2}\gamma _1 \gamma _2 \xi _{ \pm 1/2}^1 
  =  \pm \,\frac{1}{2}\xi _{ \pm 1/2}^1 \,  ,
\lbl{68} \ee
\be
 - \frac{i}{2}\gamma _1 \gamma _2 \xi _{ \pm 1/2}^2  
 =  \pm \,\frac{1}{2}\xi _{ \pm 1/2}^2 \, .
\lbl{69} \ee
The states (\ref{66}) have the property that under
the $\vartheta=\pi$ rotation, the spin 1/2 state $\xi_{1/2}^1$ transforms into the
spin $-1/2$ state $\xi_{-1/2}^1$, and vice versa. Analogous hold for the other set
of states, $\xi_{1/2}^2$, $\xi_{-1/2}^2$.

Let us stress again that
the transformation in the above example is of the type $\Phi' = R \Phi R^{-1}$.
This is a reason that, under such a transformation, a spinor of one ideal
is transformed into the spinor of a
different ideal. A transformation $R^{-1}$, acting from the right, mixes
the ideals. Another kind of transformation is $\Phi' = R \Phi$, in which case
there is no mixing of ideals. Such are the usual transformations of spinors.
By considering the objects of the entire Clifford algebra and possible transformations
among them, we find out that spinors are not a sort of objects that transform
differently than vectors under rotations. They can transform under rotations in the same
way as vectors, i.e., according to $\Phi' = R \Phi R^{-1}$. Here $\Phi$ can be a vector,
spinor or any other object of Clifford algebra. In addition to this kind of transformations,
there exist also the other kind of transformations, namely,
$\Phi' = R \Phi$, where again $\Phi$ can be any object of $Cl(1,3)$, including a
vector or a spinor. These are particular cases of the more general transformations,
$\Phi' = R \Phi S$, considered in Sec.\,1.

\subsection{Space inversion}

Let us now consider space inversion, under which the basis vectors of a reference frame
transform according to
\be
\gamma _0  \to \gamma '_0  = \gamma _0 \,,\,\,\,\,\,\gamma _r  \to \gamma '_r 
 =  - \gamma _r \,,\,\,\,r = 1,2,3 \, .
\lbl{70} \ee

The vectors of the Witt basis (\ref{16}) then transform as
\be
\begin{array}{l}
 \theta _1  \to \frac{1}{2}(\gamma _0  - \gamma _3 ) = \bar \theta _1 \, ,\,\, \\ 
 \theta _2  \to \frac{1}{2}( - \gamma _1  - i\gamma _2 ) =  - \theta _2 ,  \\ 
 \bar \theta _1  \to \frac{1}{2}(\gamma _0  + \gamma _3 ) = \theta _1 \, , \\ 
 \bar \theta _2  \to \frac{1}{2}( - \gamma _1  + i\gamma _2 ) 
 =  - \bar \theta _2 \, . \\ 
 \end{array}'
\lbl{71} \ee
A spinor of the first left ideal transforms as\ci{PavsicSpinorInverse}
\be
(\psi ^{11} {\ul 1}  + \psi ^{21} \theta _1 \theta _2  + \psi ^{31} \theta _1 
 + \psi ^{41} \theta _2 )\,\bar \theta _1 \bar \theta _2 \,\, 
 \to ( - \psi ^{11} {\ul 1} + \psi ^{21} \bar \theta _1 \theta _2 \,
  - \psi ^{31} \bar \theta _1 
  + \psi ^{41} \theta _2 )\,\theta _1 \bar \theta _2 \, .
\lbl{72} \ee
The latter equation shows that a left handed spinor of the first ideal
transforms into a right handed spinor of the third ideal.

In general, under space inversion, the matrix of the spinor basis elements
\be
\xi _{\alpha i}  = \left( {\begin{array}{*{20}c}
   {f_1 } & {f_2 } & {\bar \theta _1 f_3 } & {\theta _1 f_4 }  \\
   {\theta _1 \theta _2 f_1 } & {\bar \theta _1 \bar \theta _2 f_2 } 
   & {\theta _2 f_3 }
    & {\bar \theta _2 f_4 }  \\
   {\theta _1 f_1 } & {\bar \theta _1 f_2 } & {f_3 } & {f_4 }  \\
   {\theta _2 f_1 } & {\bar \theta _2 f_2 } & {\bar \theta _1 \theta _2 f_3 } 
   & {\theta _1 \bar \theta _2 f_4 }  \\
\end{array}} \right) ,
\lbl{73} \ee
transforms into
\be
\xi '_{\alpha i}  = \left( {\begin{array}{*{20}c}
   { - f_3 } & { - f_4 } & { - \theta _1 f_1 } & { - \bar \theta _1 f_2 }  \\
   {\bar \theta _1 \theta _2 f_3 } & {\theta _1 \bar \theta _2 f_4 } 
   & {\theta _2 f_1 } & {\bar \theta _2 f_2 }  \\
   { - \bar \theta _1 f_3 } & { - \theta _1 f_4 } & { - f_1 } & { - f_2 }  \\
   {\theta _2 f_3 } & {\bar \theta _2 f_4 } & {\theta _1 \theta _2 f_1 }
    & {\bar \theta _1 \bar \theta _2 f_2 }  \\
\end{array}} \right) .
\lbl{74} \ee
The matrix of components
\be
\psi ^{\alpha i}  = \left( {\begin{array}{*{20}c}
   {\psi ^{11} } & {\psi ^{12} } & {\psi ^{13} } & {\psi ^{14} }  \\
   {\psi ^{21} } & {\psi ^{22} } & {\psi ^{23} } & {\psi ^{24} }  \\
   {\psi ^{31} } & {\psi ^{32} } & {\psi ^{33} } & {\psi ^{34} }  \\
   {\psi ^{41} } & {\psi ^{42} } & {\psi ^{43} } & {\psi ^{44} }  \\
\end{array}} \right)~~
{\rm transform~ into}~~
\psi ^{\alpha i}  = \left( {\begin{array}{*{20}c}
   { - \psi ^{33} } & { - \psi ^{34} } & { - \psi ^{31} } & { - \psi ^{32} }  \\
   {\psi ^{43} } & {\psi ^{44} } & {\psi ^{41} } & {\psi ^{42} }  \\
   { - \psi ^{13} } & { - \psi ^{14} } & { - \psi ^{11} } & { - \psi ^{12} }  \\
   {\psi ^{23} } & {\psi ^{24} } & {\psi ^{21} } & {\psi ^{22} }  \\
\end{array}} \right) .
\lbl{76} \ee
By comparing (\ref{73}) and (\ref{74}), or by inspecting (\ref{76}), we find that 
the spinor of the 1st  ideal transforms into the spinor of the 3rd ideal, and the
spinor of the 2nd ideal transforms into the spinorof the 4th ideal.

\section{Generalized Dirac equation (Dirac-K\"ahler equation)}

Let us now consider the Clifford algebra valued fields, $\Phi(x)$, that depend
on position $x\equiv  x^\mu$ in spacetime. We will assume that a field $\Phi$
satisfies the following equation\,\ci{DiracKaehler}
(see also refs.\,\ci{PavsicSpinorInverse,PavsicOrthoSympl}):
\be
(i\,\gamma ^\mu  \partial _\mu   - m)\Phi  = 0, ~~~~
\Phi  = \phi^A \gamma _A  = \psi ^{\tilde A} \xi _{\tilde A} 
 = \psi ^{\alpha i} \xi _{\alpha i} \, .
\lbl{77} \ee
where $\gam_A$ is a multivector basis of $Cl(1,3)$, and
$\xi_{\tl A} \equiv \xi_{\alpha i}$ is a spinor basis of $Cl(1,3)$,
or more precisely, of its complexified version if $\psi^{\alpha i}$ are complex-valued.
Here $\alpha$  is the spinor index of a left minimal ideal, whereas the
$i$   runs over four left ideals of $Cl(1,3)$.

Multiplying Eq.\,(\ref{77}) from the left by $(\xi ^{\tilde A} )^\ddag$,
where $\ddag$ is the operation of reversion that reverses the order of
vectors in a product, and using the relation
\be
\langle (\xi ^{\tilde A} )^\ddag  \gamma ^\mu  \xi _{\tilde B} \rangle _S  
\equiv {(\gamma ^\mu  )^{\tilde A}} _{\tilde B} \, ,
\lbl{78} \ee
and where $\langle ~~\rangle_S$ is the (properly
normalized\,\ci{PavsicKaluzaLong}) scalar part of
an expression, we obtain the following matrix form of the equation (\ref{77}):
\be
\left( {i\,{(\gamma ^\mu  )^{\tilde A}} _{\tilde B}\, \partial _\mu 
- m \,{\delta^{\tilde A}}_{\tilde B}} \right)\psi ^{\tilde B}  = 0 \, .
\lbl{79} \ee
The $16 \times 16$ matrices can be factorized according to
\be
{(\gamma ^\mu  )^{\tilde A}}_{\tilde B} 
 = {(\gamma ^\mu)^\alpha}_\beta \, {\delta^i}_j \, ,
\lbl{80} \ee
where ${(\gamma^\mu)^\alpha}_\beta$ are $4 \times 4$ Dirac matrices.
Using the latter relation (\ref{80}), we can write Eq.\,(\ref{79}) as
\be
\left( {i\,{(\gamma^\mu)^\alpha}_\beta \, \partial_\mu 
- m \, {\delta^\alpha}_\beta} \right)\psi ^{\beta i}  = 0 \, ,
\lbl{81} \ee
or more simply,
\be
(i\,\gamma ^\mu  \partial _\mu   - m)\psi ^i  = 0 \, .
\lbl{82} \ee
In the last equation we have omitted the spinor index $\alpha$.

The action that leads to the generalized Dirac equation (\ref{77}) is
\be
I = \int {{\rm{d}}^4 x\,\,} \bar \psi ^i 
(i\,\gamma ^\mu  \partial _\mu   - m)\psi ^j z_{ij} \, .
\lbl{83} \ee
This is an action that describes four spinors $\psi^i$, belonging to the four minimal left
ideals of $Cl(1,3)$. Here $z_{ij}$ is the metric in the space of ideals. It is a part
of the metric
\be 
(\xi _{\tilde A} )^\ddag  *\xi _{\tilde B}  = z_{\tilde A\tilde B}  
= z_{(\alpha i)(\beta j)}  = z_{\alpha \beta } z_{ij}
\lbl{83a}
\ee 
of the Clifford algebra $Cl(1,3)$, represented in the
basis $\xi_{\tl A}$:
\be
 \,\,\,z_{ij}  = \left( \begin{array}{l}
 1\,\,\,\,\,\,\,0\,\,\,\,\,\,\,\,0\,\,\,\,\,\,0 \\ 
 0\,\,\,\,\,\,\,1\,\,\,\,\,\,\,\,0\,\,\,\,\,\,0 \\ 
 0\,\,\,\,\,\,0\,\,\,\,\, - 1\,\,\,\,\,\,0 \\ 
 0\,\,\,\,\,\,0\,\,\,\,\,\,\,\,0\,\,\, - 1 \\ 
 \end{array} \right)\,\,,\,\,\,\,\,\,\,\,\,\,\,\,\,
 z_{\alpha \beta }  = \left( \begin{array}{l}
 1\,\,\,\,\,\,\,0\,\,\,\,\,\,\,\,0\,\,\,\,\,\,0 \\ 
 0\,\,\,\,\,\,\,1\,\,\,\,\,\,\,\,0\,\,\,\,\,\,0 \\ 
 0\,\,\,\,\,\,0\,\,\,\,\, - 1\,\,\,\,\,\,0 \\ 
 0\,\,\,\,\,\,0\,\,\,\,\,\,\,\,0\,\,\, - 1 \\ 
 \end{array} \right) . 
\lbl{84} \ee

Gauge covariant action is
\be
\,I = \int {{\rm{d}}^4 x\,\,} \bar \psi ^i (i\,\gamma ^\mu  D_\mu   - m)\psi ^j z_{ij}~,
~~~~~~D_\mu  \psi^i  = \partial_\mu  \psi^i  + {{G_\mu}^i}_j \psi^j \, .
\lbl{85} \ee
This action contains the ordinary particles and mirror particles.
The first and the second columns of the matrix $\psi^{\alpha i}$, written explicitly
in eq.\,(\ref{37}) describe the ordinary particles, whereas the third and the forth column in
(\ref{37}) describe {\it mirror particles}.

The SU(2) gauge group acting within the 1st and 2nd 
ideal can be interpreted as the weak interaction gauge
group for ordinary particles. The  SU(2) gauge group
acting within the 3rd and 4th  ideal can be interpreted as
the weak interaction gauge group for mirror particles.
The corresponding two kinds of weak interaction gauge fields that can be
transformed into each other by space inversion are contained in 
${{G_\mu}^i}_j$, which is a generalized gauge field occurring in the
covariant action (\ref{85}).

Mirror particles were first proposed by  Lee and Yang\,\ci{LeeYang}.
Subsequently, the idea of mirror particles has been pursued by Kobzarev et al.\,
\ci{Kobzarev}, and in Refs.\,\ci{PavsicMirror}--\ci{Foot4}.
The possibility that mirror particles are responsible for dark matter
has been explored in many works, e.g., in \ci{Hodges}--\ci{Ciarcelluti3}.
A demonstration that mirror particles can be explained in terms of
algebraic spinors (elements of Clifford algebras) was presented in
Ref.\,\ci{PavsicSpinorInverse}.

\vs{3mm}

\centerline{\bf Acknowledgment}

This work has been supported by the Slovenian Research Agency.

\end{document}